\documentclass[11pt,twoside]{article}
\usepackage{natbib}
\usepackage{asp2010}

\resetcounters

\begin{document}

\title{Scattering polarization and the Hanle effect in H$\alpha$ as a probe of chromospheric 
magnetism: Modeling vs. Observations}
\author{J.~\v{S}t\v{e}p\'an$^{1,2}$, J.~Trujillo~Bueno$^{1,2,3}$, R.~Ramelli$^4$ and M.~Bianda$^4$
\affil{$^1$Instituto de Astrof\'{\i}sica de Canarias, E-38205 La Laguna, Tenerife, Spain}
\affil{$^2$Departamento de Astrof\'\i sica, Universidad de La Laguna, Tenerife, Spain}
\affil{$^3$Consejo Superior de Investigaciones Cient\'\i ficas, Spain}
\affil{$^4$Istituto Ricerche Solari Locarno, CH-6605 Locarno Monti, Switzerland}}

\begin{abstract}
The Hanle effect in strong spectral lines is the physical mechanism that should be increasingly exploited for quantitative explorations of the magnetism of the quiet solar chromospheric plasma. Here we show, by means of multilevel radiative transfer calculations and new spectropolarimetric observations, that the 
amplitude and shape of the scattering polarization profiles of the H$\alpha$ line is very sensitive to the strength and structure of the chromospheric magnetic field. The interpretation of the observations in terms of one-dimensional radiative transfer modeling suggests that there is an abrupt magnetization in the upper chromosphere of the quiet Sun. 
\end{abstract}


\section{Introduction}

Spectroscopy of the hydrogen H$\alpha$ line at $6562.8\,\AA$ has played a crucial role to visualize that 
the ``quiet" solar chromosphere is a fibrilar dominated-magnetism medium \citep[e.g., the review by][]{rutten07}, but quantitative information on the magnetic field vector can only be obtained through spectropolarimetry. 

Unfortunately, the spectral line polarization produced by the Zeeman effect is of little practical interest here because the magnetic splitting of the $\sigma$ and $\pi$ components of the line transitions are practically negligible compared with the very significant Doppler width of the H$\alpha$ line, so that in the quiet Sun the Zeeman effect only produces observable  circular polarization signals of very low amplitude (with $V/I{\sim}0.01\%$ when using today's telescopes), which nonetheless are practically insensitive to the physical conditions of the true chromosphere 
\citep[see the review by][]{jtb10sacpeak}. 

Fortunately, as shown by \citet{stepanjtb10asym}, the amplitude and shape of the H$\alpha$ 
linear polarization profiles produced by scattering processes in the solar atmosphere are very sensitive (through the Hanle effect) to the strength and structure of the magnetic field in the upper chromosphere of the quiet Sun. Their conclusion was based on radiative transfer modeling of the peculiar line core asymmetry (LCA) of the 
$Q/I$ profile observed by \citet{gandorfer00} in a quiet region close to the solar limb, an asymmetry which is however not seen in the observed $I(\lambda)$ profile. The main aim of this paper is to clarify further why we believe that the measurement and interpretation of the $Q/I$ and $U/I$ profiles produced by scattering processes and the Hanle effect in H$\alpha$ might lead to an important breakthrough in our empirical understanding of chromospheric magnetism. To this end, here we show both additional radiative transfer calculations and an example of the 
spectropolarimetric observations we are obtaining with the Z\"urich Imaging Polarimeter (ZIMPOL) attached to the Gregory Coud\'e Telescope (GCT) of the Istituto Ricerche Solari Locarno (IRSOL). Interestingly, these new observations show non-zero signals in both $Q/I$ and $U/I$.


\section{Basic information about the Hanle effect in H${\alpha}$}

\begin{figure}
\begin{tabular}{cc}
\includegraphics[width=2.5in]{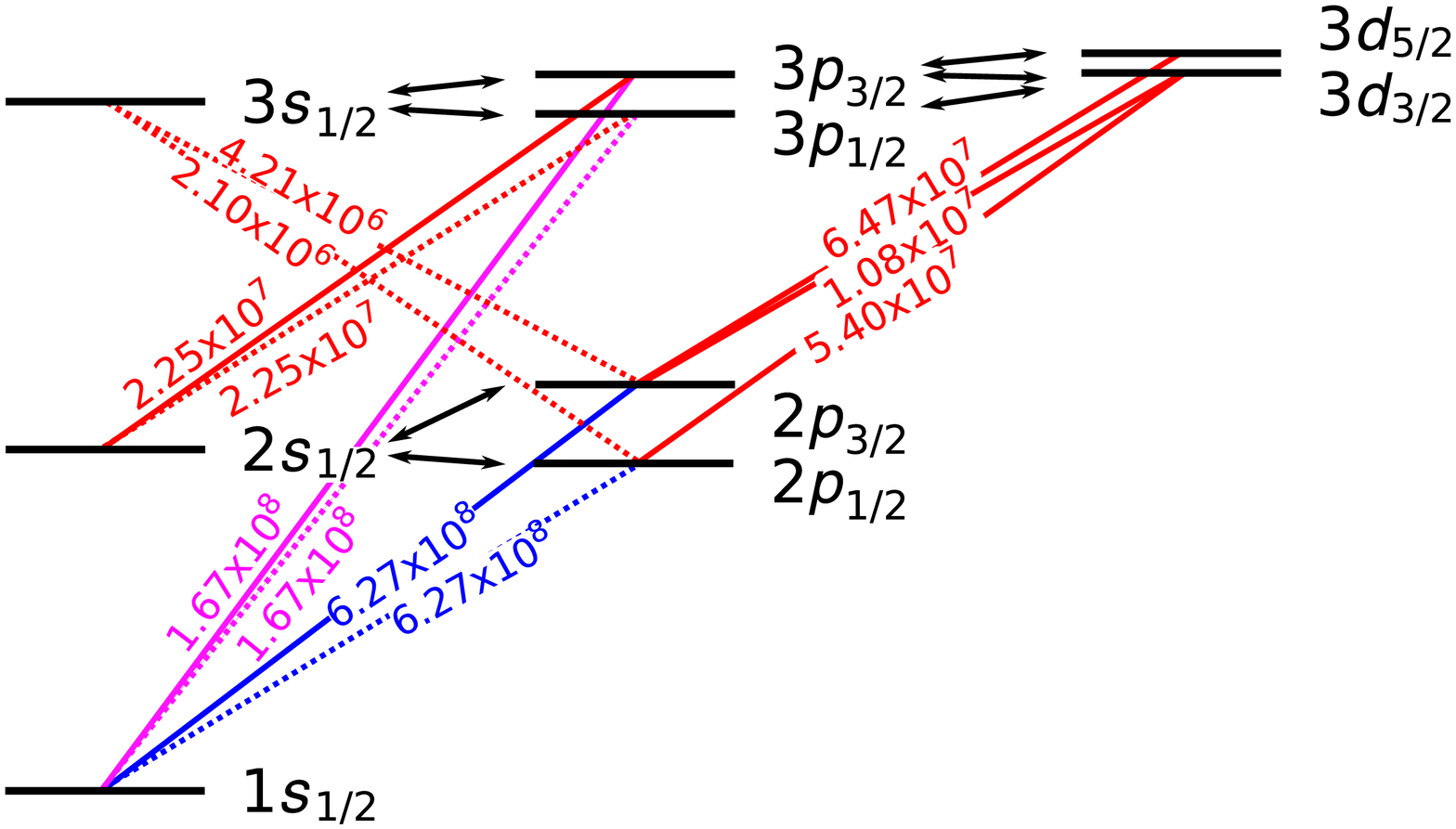} & \includegraphics[width=2.5in]{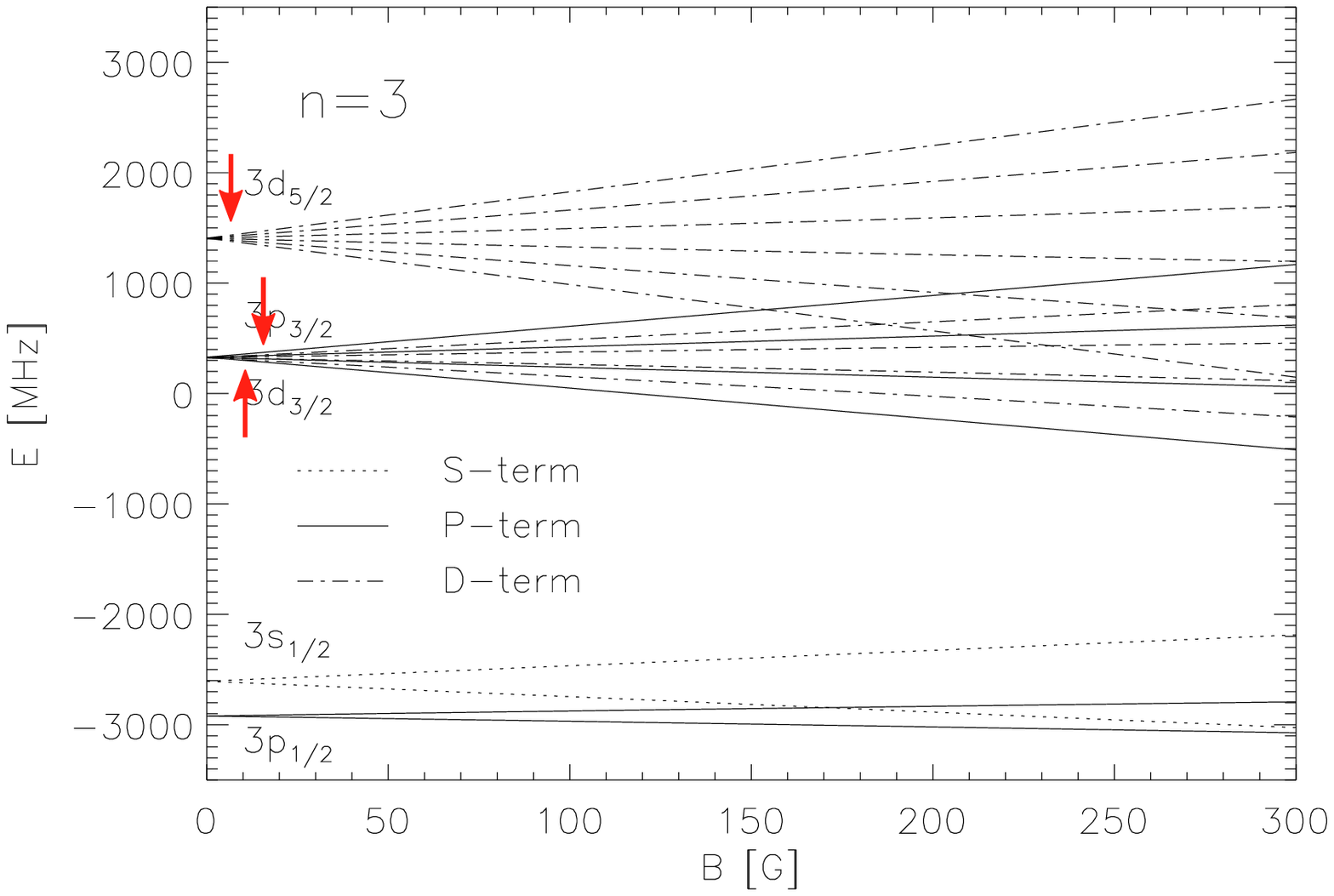}
\end{tabular}
\caption{Left panel: Grotrian diagram of neutral hydrogen showing H$\alpha$ (red lines), Ly$\alpha$ (blue lines), and Ly$\beta$ (violet lines). The four red solid lines correspond to the transitions which make a significant contribution to the solar H$\alpha$ polarization. Right panel: Zeeman splittings of the FS levels of $n=3$ calculated in the incomplete Paschen-Back effect regime. The red arrows indicate the $B_H$ values of the individual levels.}
\label{stepan:fig:ha}
\end{figure}

The physical origin of the so-called scattering line polarization is nothing but the selective emission and absorption processes produced by the presence of atomic alignment in the line's levels, which results from radiative transitions produced by anisotropic radiation \citep[e.g.,][]{jtb01}.

The H$\alpha$ line is produced by the $3\ell j$--$2\ell'j'$ transitions of neutral hydrogen (see the left panel of Fig.~\ref{stepan:fig:ha}). The Zeeman splitting of the upper levels of H$\alpha$ is shown in the right panel of the same figure. 
The selection rule $\Delta{l}={\pm}1$ for radiative transitions inhibits radiative couplings between levels ($n,\,l,\, j$) and ($n,\,l{\pm}1,\, {j}'$). Therefore, for $B{\lesssim}100$ G we can safely neglect quantum coherences between different $j$-levels. The thermal broadening of the H$\alpha$ line is large due to the high chromospheric temperatures and low atomic mass of hydrogen. A typical Doppler width is about 0.3\,$\AA$. For magnetic strengths $B{\lesssim}100$ G the wavelength shifts of the Zeeman $\sigma$ and $\pi$ components do not produce any noticeable contribution to the Stokes $Q$ and $U$ profiles, which are dominated by the presence of atomic level alignment and its modification by the Hanle effect. Since the separation of the fine structure (FS) components of the H${\alpha}$ line is about 0.1\,$\AA$, the ensuing transitions do overlap. 

Via the Hanle effect each of the FS levels is sensitive to a different strength of the magnetic field according to the expression $B_H=1.137\times 10^{-7}/t_{\rm life}g\,,$ where $B_H$ (in gauss) is the so-called critical Hanle field of the level under consideration, $t_{\rm life}$ its lifetime in seconds, and $g$ its Land\'e factor. The linear polarization of the scattered radiation is mostly sensitive to the Hanle effect of magnetic fields with intensities of the order of $B_H$. If the magnetic field is weaker than about $0.1\,B_H$ the polarization state of the scattered radiation is not modified.  If $B>100\,B_H$, the polarization state is sensitive only to the field direction (i.e., the so-called Hanle saturation regime). The critical fields of the upper levels of H$\alpha$ range from approximately 6 to about 15\,G. Note also that only the FS levels with angular momentum $j>1/2$ may carry atomic alignment and thus contribute to the scattering polarization of the H$\alpha$ line.


\section{Radiative Transfer Calculations}

Our radiative transfer calculations of the H$\alpha$ scattering polarization profiles and their modification due to the action of the Hanle effect are carried out within the framework of the density-matrix theory of spectral line polarization, which uses the approximation of complete frequency redistribution \citep[see][]{ll04}. The numerical solution of this type of problem requires finding, at each spatial grid point of the model atmosphere under consideration, and for each $j$-level of the chosen atomic model, the self-consistent values of the multipolar components of the density matrix. We achieve this through the application of efficient iterative methods and formal solvers of the Stokes-vector transfer equations \citep[see][for a detailed review]{jtb03}.


\subsection{H$\alpha$ in an isothermal unmagnetized atmosphere}

\begin{figure}
\begin{tabular}{cc}
\includegraphics[width=2.5in]{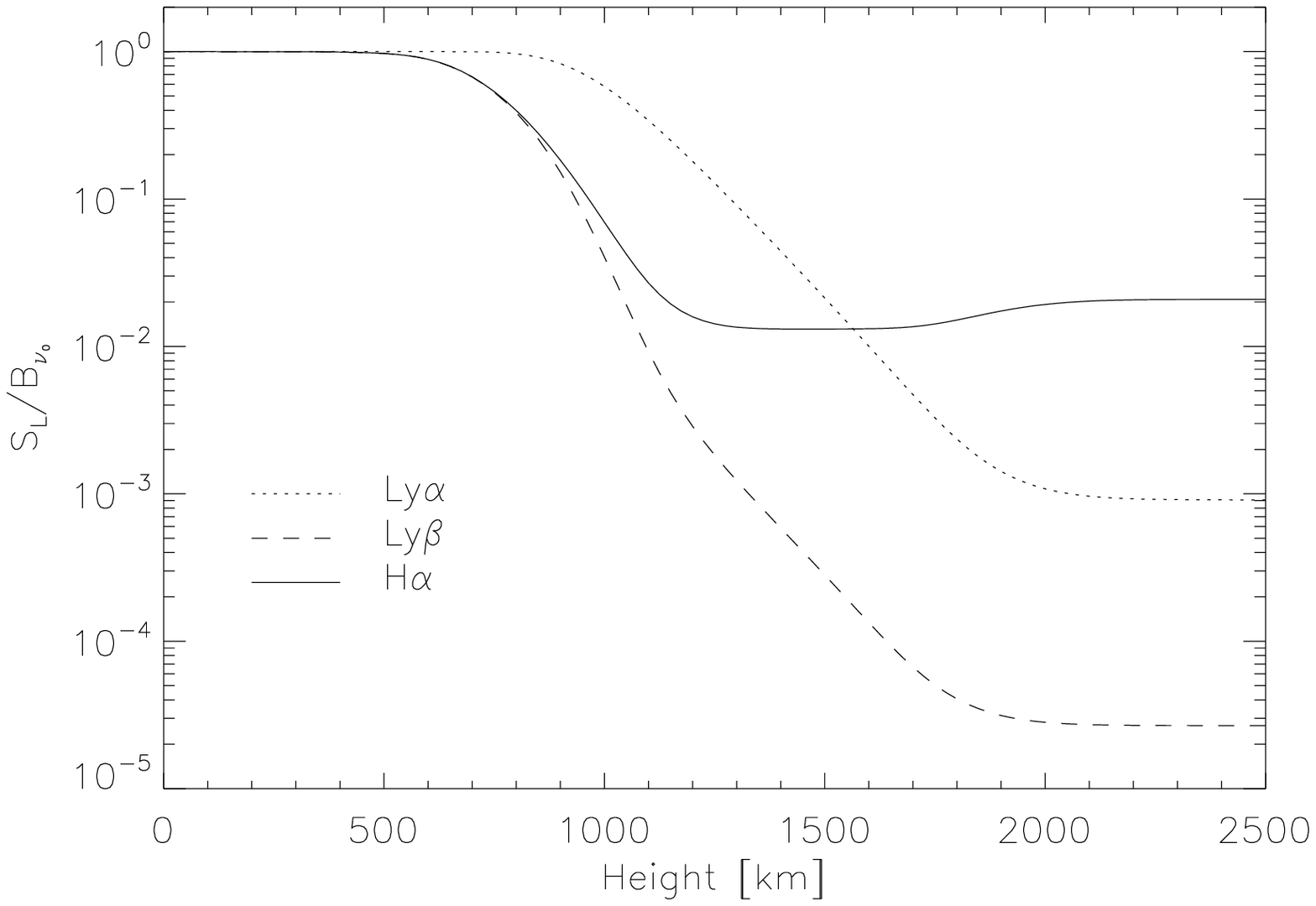} & \includegraphics[width=2.5in]{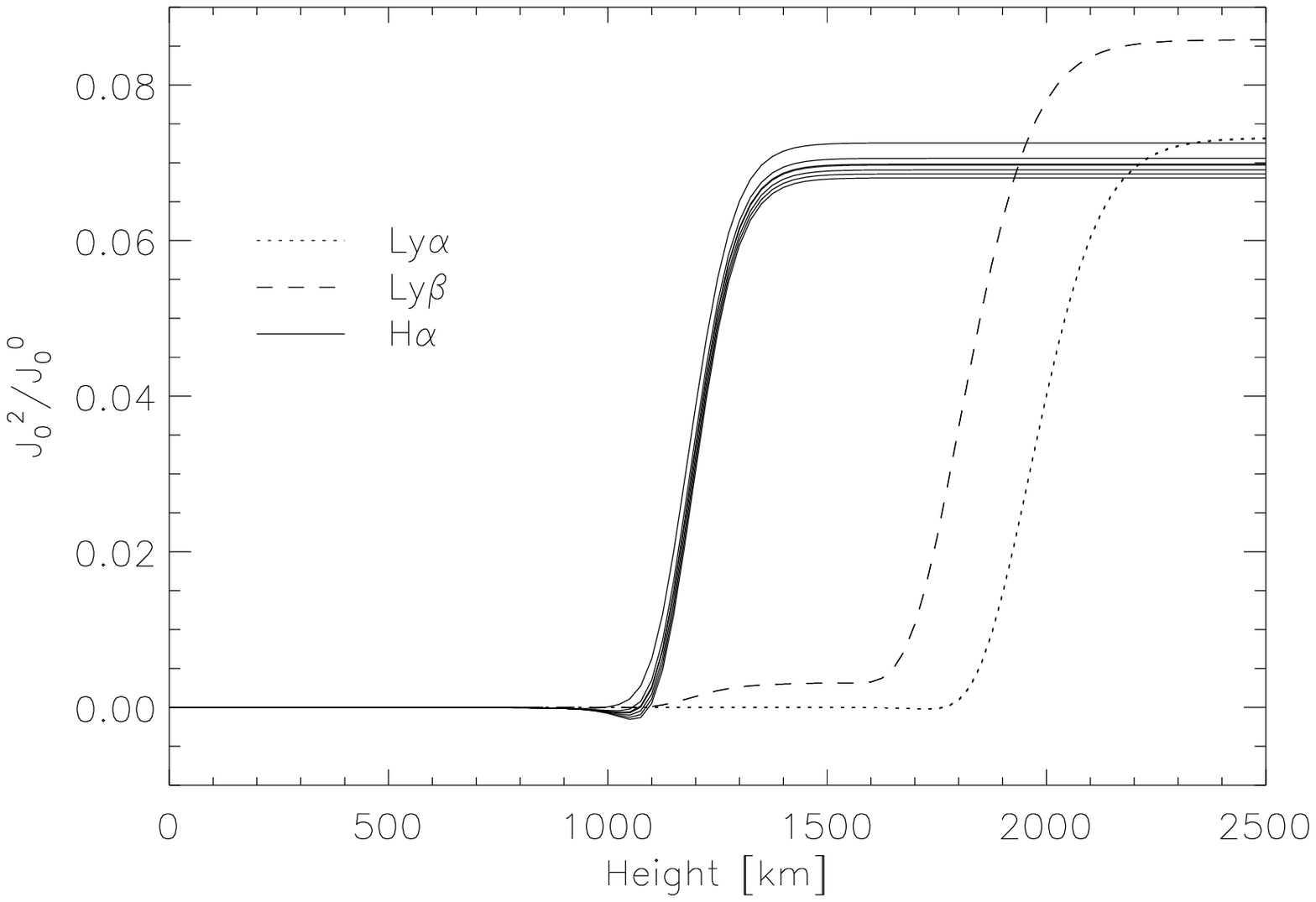}
\end{tabular}
\caption{Left panel: Line source functions in a semi-infinite isothermal atmosphere with $n_e=n_p=4\times 10^{10}\,{\rm cm^{-3}}$ and $T=10^4$\,K. Right panel: The corresponding fractional anisotropy, $\bar J^2_0/\bar J^0_0$, of the line radiation.}
\label{stepan:fig:sf}
\end{figure}

\begin{figure}
\begin{tabular}{cc}
\includegraphics[width=2.5in]{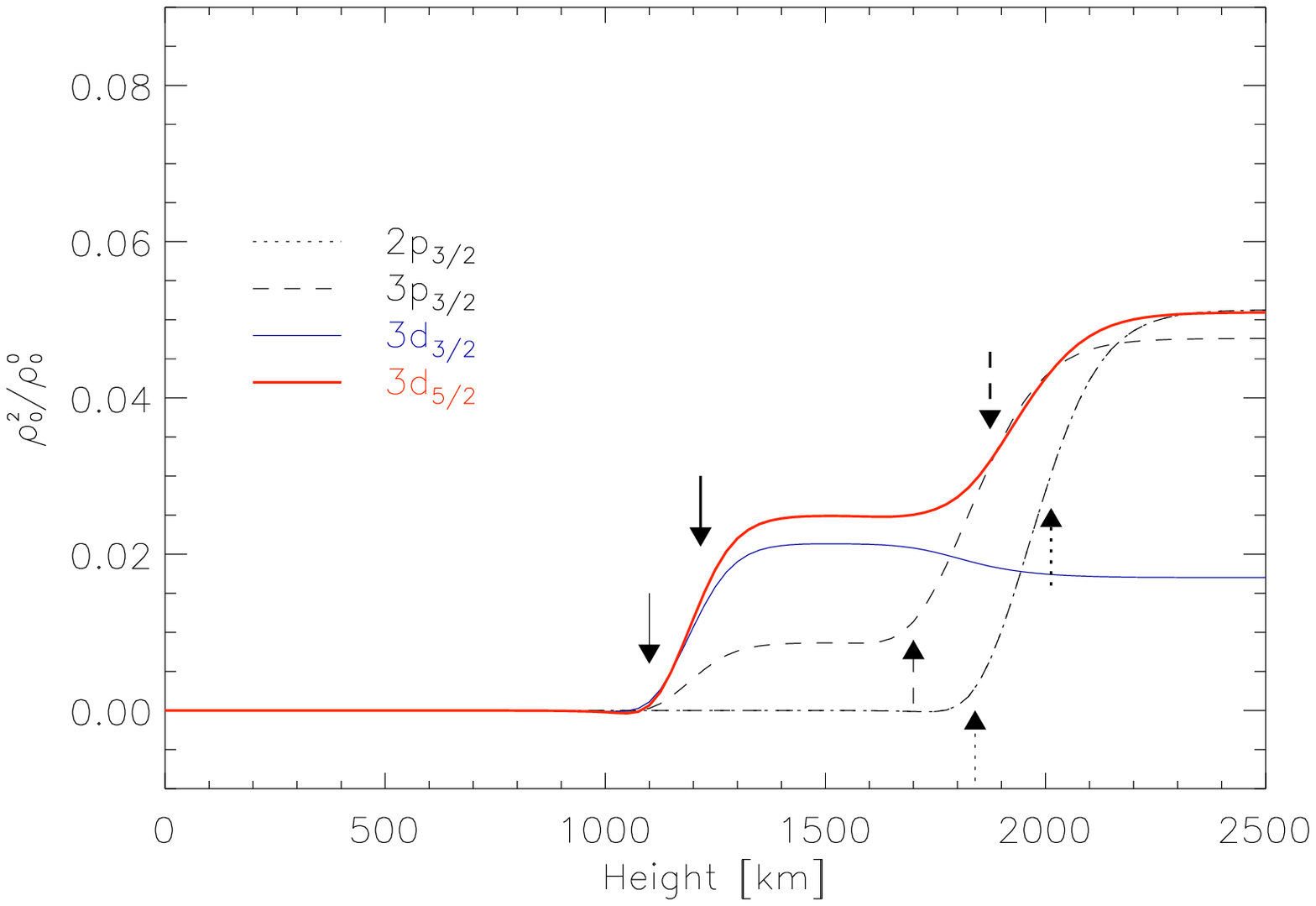} & \includegraphics[width=2.5in]{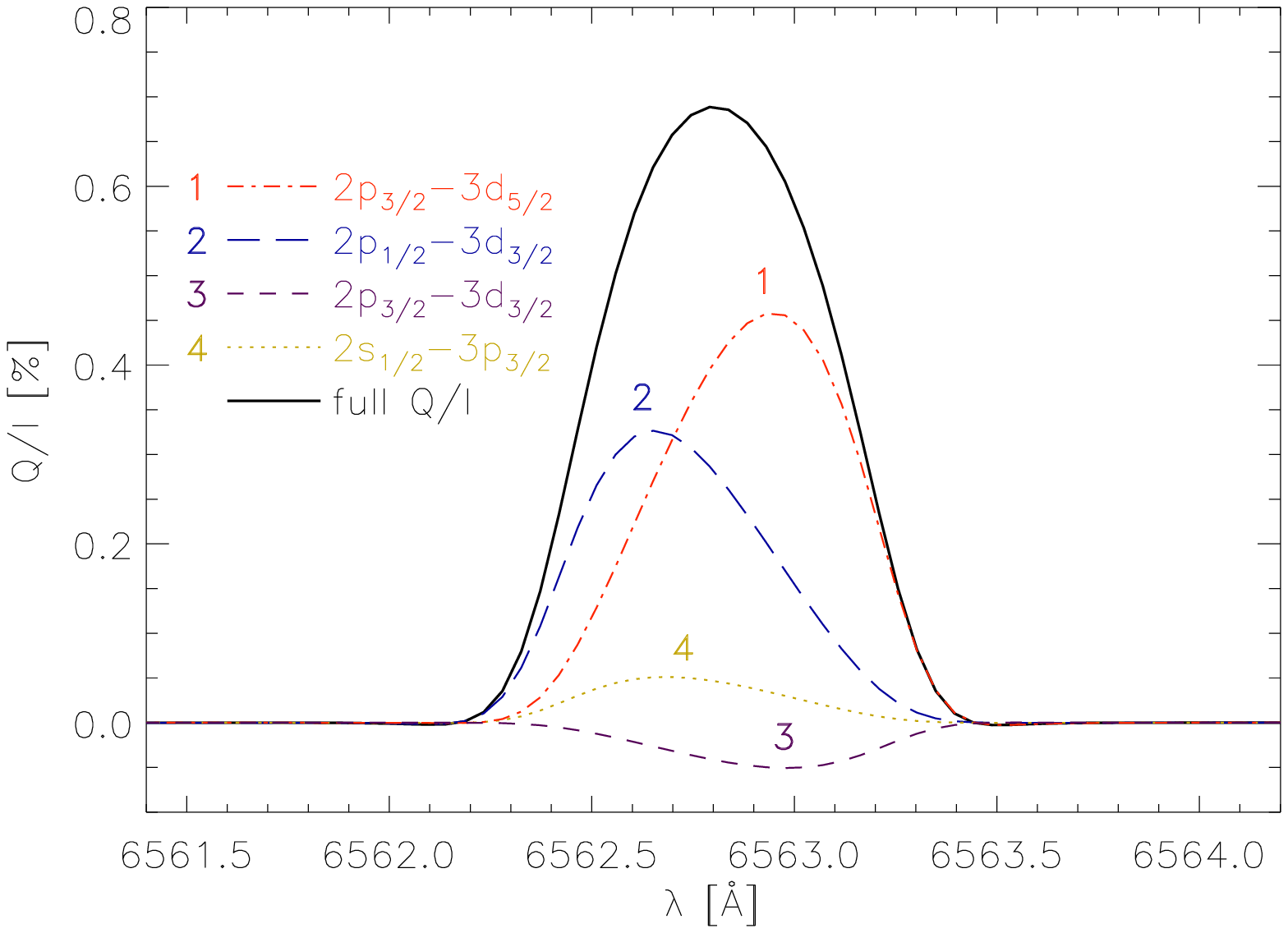}
\end{tabular}
\caption{Fractional level alignment (left panel) and the emergent fractional linear polarization profile for $\mu=0.1$ (right panel) without including any magnetic field in the isothermal atmosphere. The arrows indicate the atmospheric heights where the line optical depth is unity for line-of-sights (LOS) with $\mu=1$ and $\mu=0.1$.}
\label{stepan:fig:fp}
\end{figure}

We assume a 3-$n$-level model atom consisting of 9 FS levels and consider a semi-infinite, isothermal model atmosphere with temperature $T=10^4$\,K and constant electron and proton densities (i.e., $n_{\rm e}=n_{\rm p}=4\times 10^{10}\,{\rm cm^{-3}}$). The inelastic collisional rates of the transitions are taken from \citet{przybilla04} and the depolarizing collisional rates are calculated in the semi-classical approximation of \citet{sahal96}. We have calculated the self-consistent solution of the ensuing non-LTE scattering polarization problem. 
Fig.~\ref{stepan:fig:sf} shows the line source functions and the radiation anisotropies of the Ly$\alpha$, Ly$\beta$, and H$\alpha$ lines throughout the atmosphere.

In order to calculate the emergent linear polarization, one has to take into account the atomic polarization of the FS levels. In the left panel of Fig.~\ref{stepan:fig:fp} we show the fractional alignment of the individual levels as a function of height in the atmosphere. At the formation depths of  H$\alpha$, the strong resonant Ly$\alpha$ line is optically thick and the Ly$\alpha$ radiation is virtually isotropic there (see the right panel of Fig.~\ref{stepan:fig:sf}). Consequently, the polarization of the 2$p_{3/2}$ level is negligible and, therefore, selective absorption processes in H$\alpha$ can be neglected. The H$\alpha$ scattering polarization is fully determined by polarization of the upper levels, especially by the 3$d_j$ ones. Thanks to this fact, the emergent $Q/I$ profile can be expressed as the sum of the contributions of the individual FS transitions \citep[see the right panel of Fig.~\ref{stepan:fig:fp}; cf. ][for details]{stepanjtb10asym}.


\subsection{H$\alpha$ in an isothermal magnetized atmosphere}

\begin{figure}
\begin{tabular}{cc}
\includegraphics[width=2.5in]{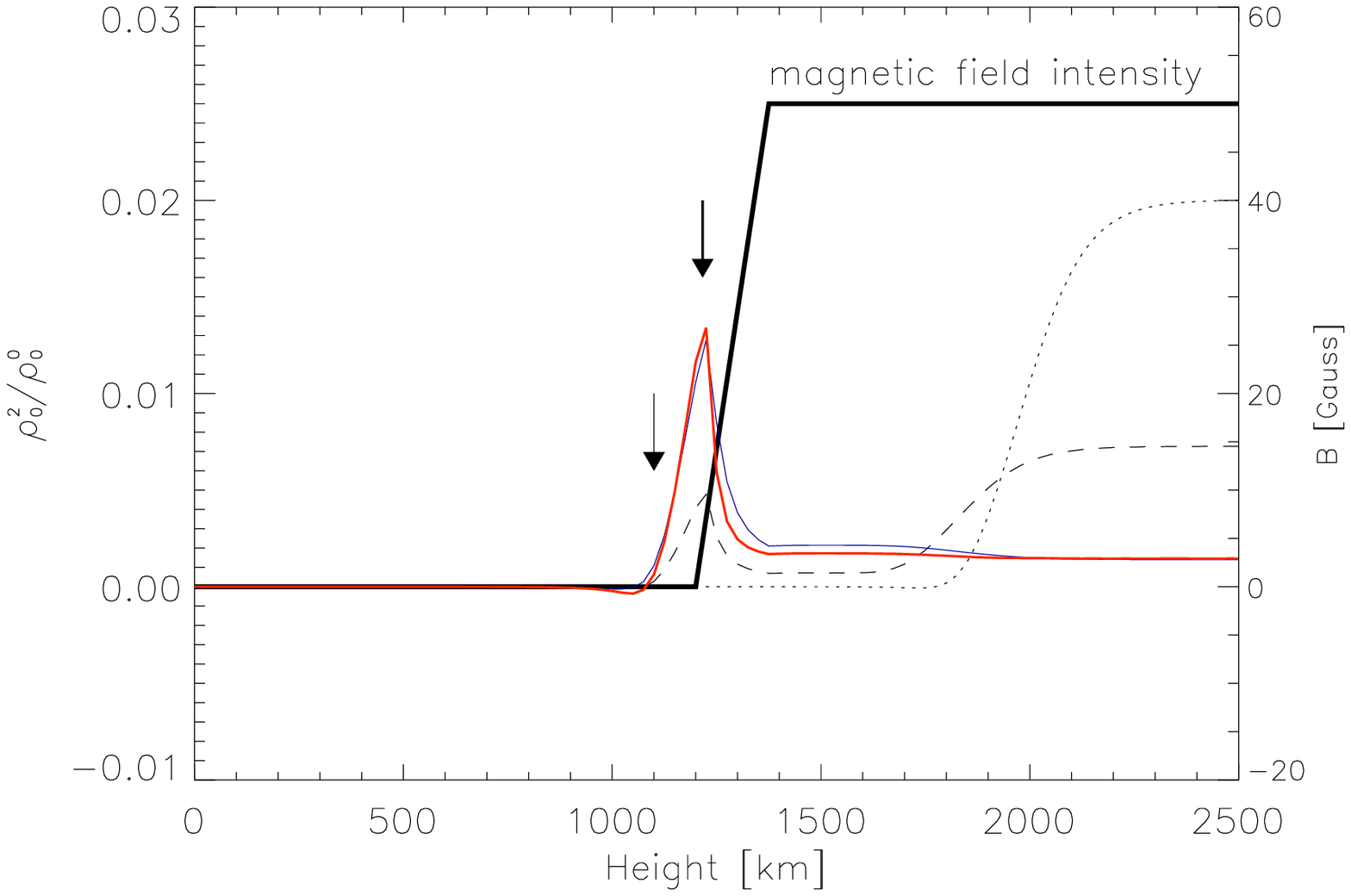} & \includegraphics[width=2.5in]{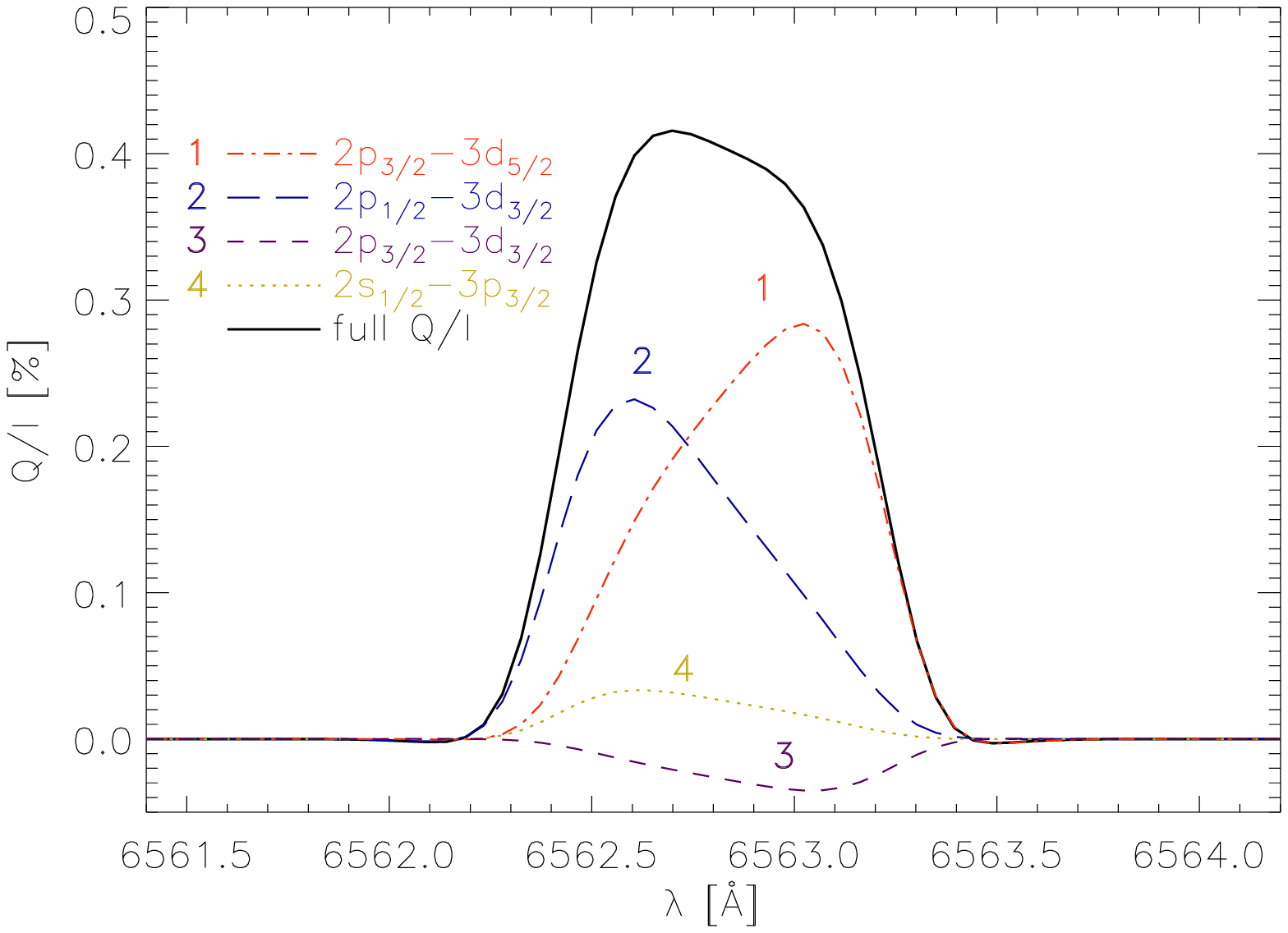}
\end{tabular}
\caption{Same as Fig.~\ref{stepan:fig:fp}, but for a magnetic field with a fixed inclination $\theta_B=65^\circ$ and with a random azimuth changing its intensity with height in the atmosphere.}
\label{stepan:fig:gr}
\end{figure}

For simplicity, here we consider the case of a magnetic field with a fixed inclination $\theta_B$ with respect to the vertical and with a random azimuth, $\chi_B$, uniformly distributed in the interval $[0,2\pi]$. Such a magnetic field does not break the cylindrical symmetry of the atmosphere and makes the numerical solution relatively simple \citep[see][for details]{stepanjtb10paper1}. If the magnetic field intensity is uniform throughout the atmosphere, the fractional level alignments $\rho^2_0/\rho^0_0$ are reduced as the field intensity increases and the amplitude of the $Q/I$ line profile is reduced accordingly \citep{stepanjtb10paper1}.

If there is a spatial gradient in the magnetic field strength, then the shape of
the resulting $Q/I$ profile may be more complicated than the typical Gaussian bell-like shape of Fig.~\ref{stepan:fig:fp}.
Such an effect is due to the different magnetic sensitivities of the (blended) H$\alpha$ line components and to the fact that different parts of the line profile originate at  slightly different depths in the model atmosphere. A clear demonstration of this effect can be achieved 
for the case in which the magnetic field intensity increases with height around the height corresponding to unit line optical depth. An academic example of such a magnetic field variation is shown in the left panel of Fig.~\ref{stepan:fig:gr}. The line center photons originate in the upper layers of the atmosphere which are strongly depolarized by the magnetic field there, and the line-center amplitude thus decreases. Given that  the different components of the line are contributing differently to the resulting $Q/I$ profile and since they are sensitive to different field strengths, a line-center asymmetry (LCA) may appear (see the right panel of Fig.~\ref{stepan:fig:gr}).


\subsection{H$\alpha$ in a semi-empirical model with magnetic field gradients}

\begin{figure}
\begin{tabular}{cc}
\includegraphics[width=2.5in]{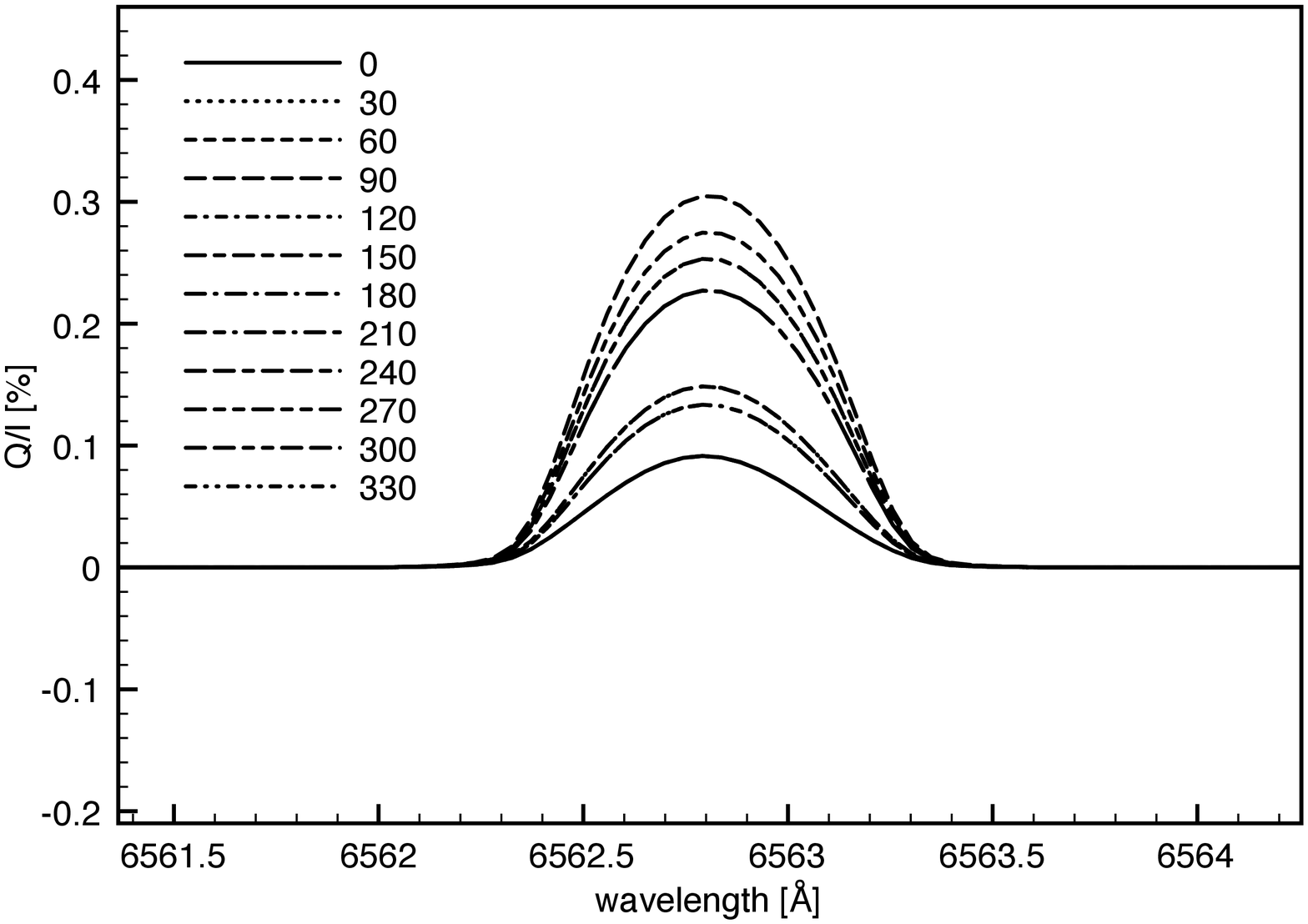} & \includegraphics[width=2.5in]{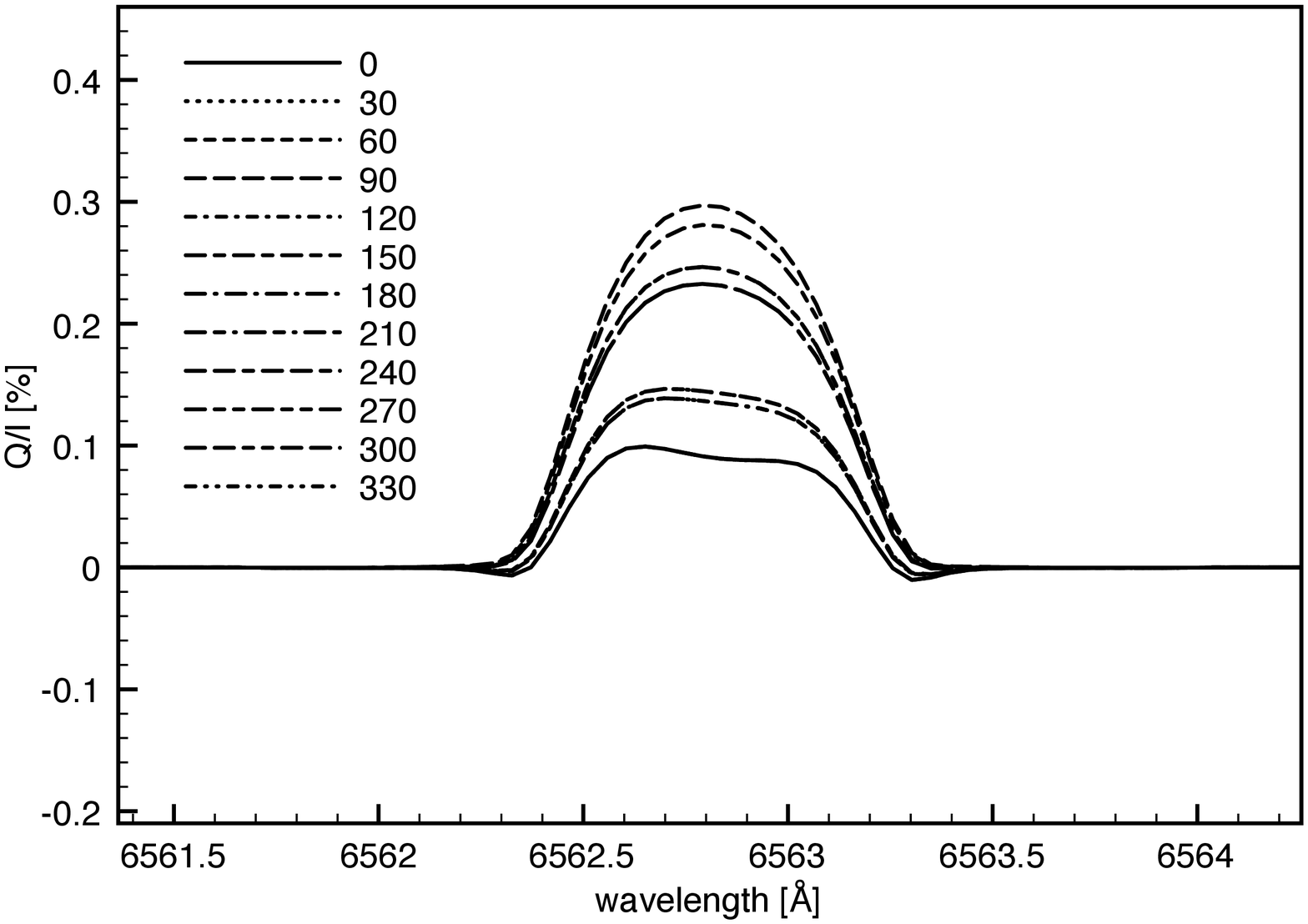}
\end{tabular}
\caption{The emergent $Q/I$ profiles at $\mu=0.1$ for different azimuths $\chi_B$ of the horizontal magnetic field (see the different line styles and the $\chi_B$ values in degrees). Left panel: Magnetic field intensity is decreasing with height around $\tau=1$. Right panel: Magnetic field intensity is increasing with height around $\tau=1$.}
\label{stepan:fig:dm}
\end{figure}

In Fig.~\ref{stepan:fig:dm} we show examples of synthetic H$\alpha$ $Q/I$ profiles obtained by assuming 
two types of magnetic field gradients in the semi-empirical model atmosphere of \citet[][hereafter, the FAL-C model]{fontenla93}. While the model with magnetic strength decreasing with height produces quite symmetric $Q/I$ profiles for all the field azimuths considered, the model with magnetic strength increasing with height leads to 
a flattening of the $Q/I$ profile and to a LCA which can be significantly stronger than when assuming a random-azimuth field, especially for $\chi_B\approx 0^\circ$ and $180^\circ$. The corresponding $U/I$ profiles do not show any noteworthy asymmetry and they are thus not plotted here. Calculations for the case of a random-azimuth field can be found in Figs. 2 and 3 of \citet{stepanjtb10asym}.

\subsection{H$\alpha$ in hydrodynamical models of chromospheric dynamics
\label{stepan:ssec:hydro}}

The solar chromosphere is a dynamical system whose time evolution is far too fast to be resolved by  spectropolarimetric observations. Consequently, only the time average of the emergent line profiles can be measured with today's telescopes. Fortunately, radiative transfer synthesis of the emergent scattering polarization  
in the available 1D hydrodynamical models of chromospheric internetwork dynamics by \citet{carlssonstein97} are a suitable step prior to performing similar calculations in 3D models of the quiet chromosphere.

\begin{figure}
\begin{tabular}{cc}
\includegraphics[width=2.5in]{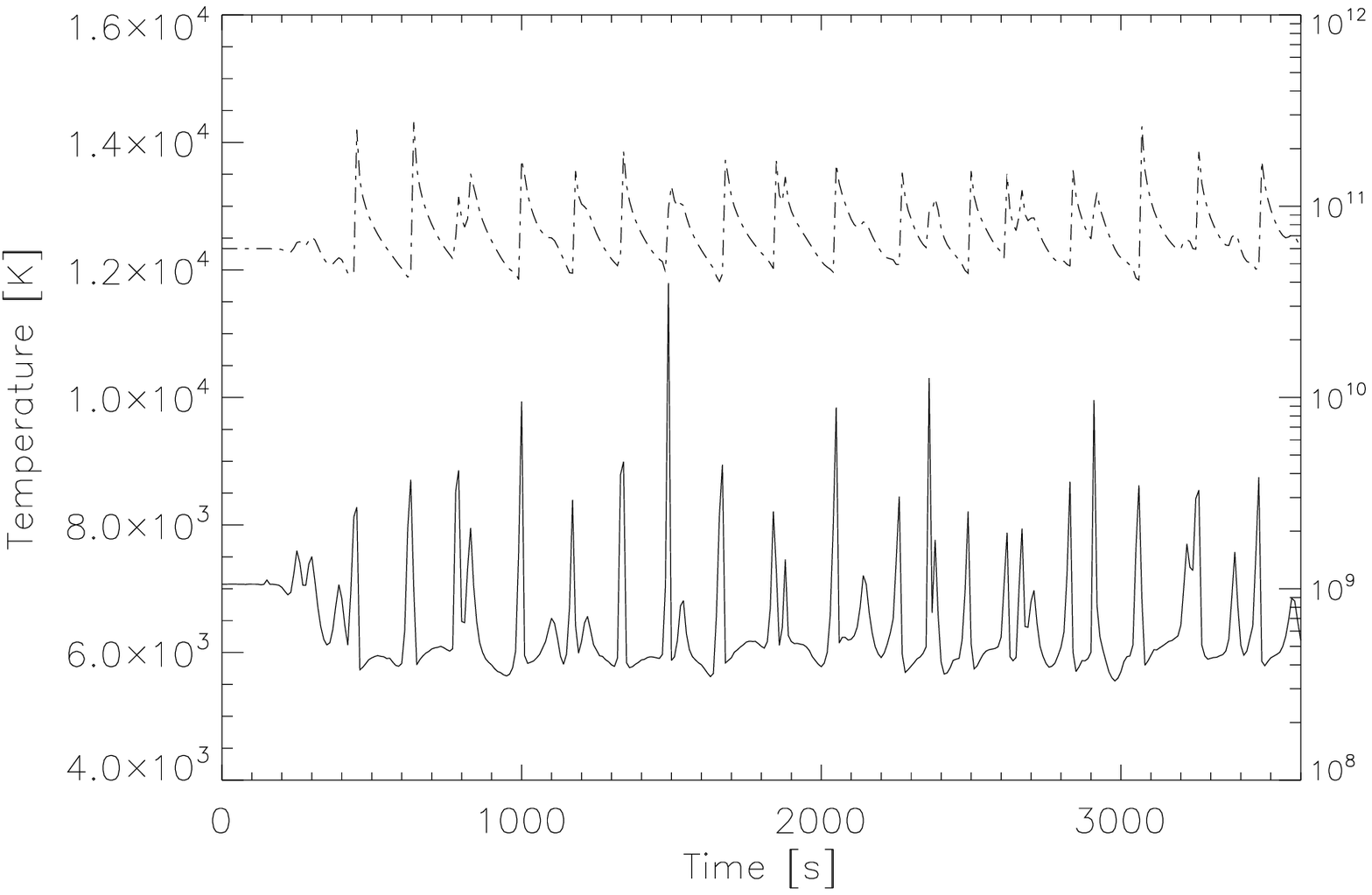} & \includegraphics[width=2.5in]{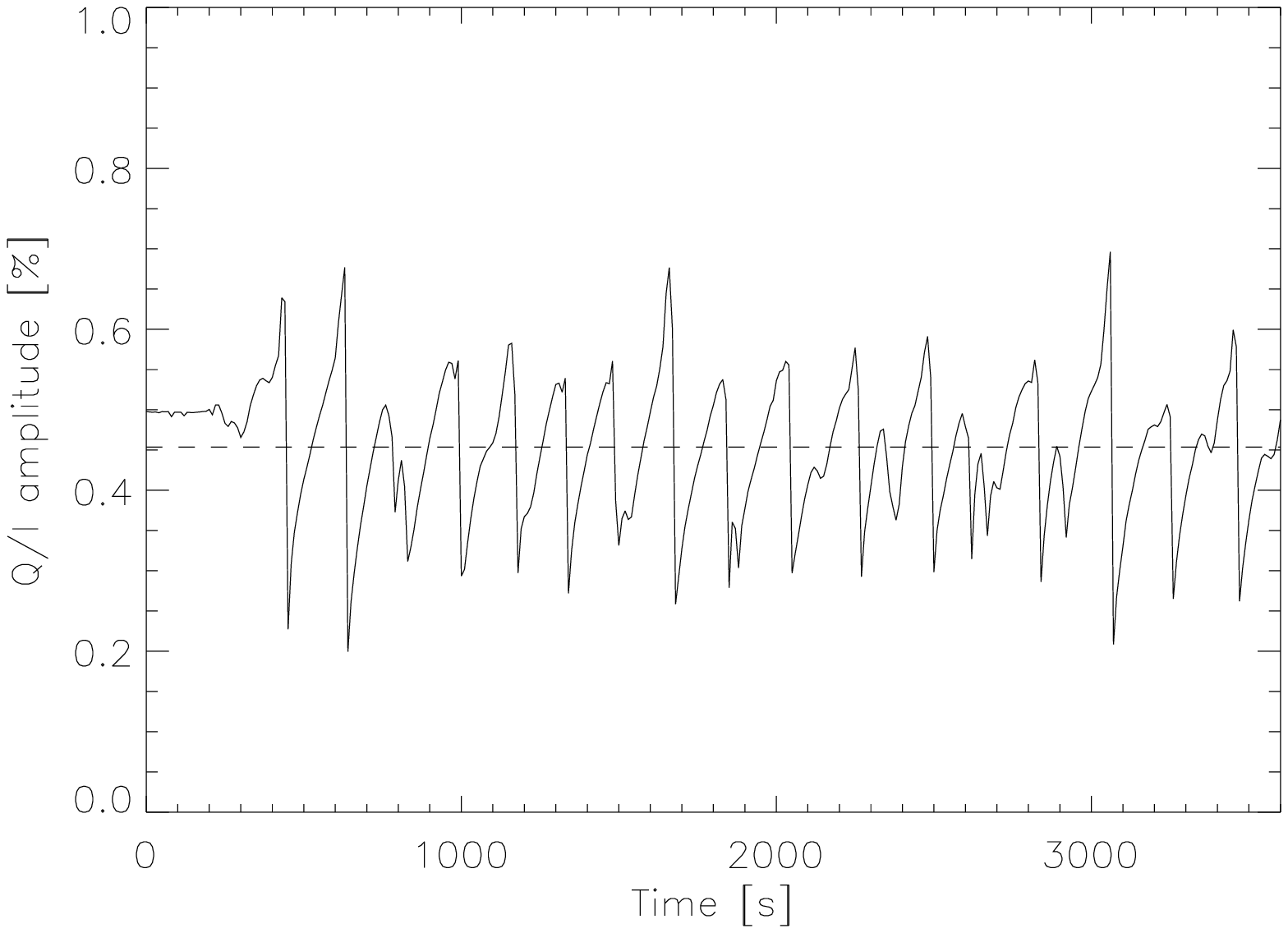}
\end{tabular}
\caption{Left panel: Time evolution of the temperature (solid line) and of the proton density (dash-dotted line) in one of hydrodynamical simulations of \citet{carlssonstein97}. Right panel: The corresponding amplitude of the H$\alpha$ fractional linear polarization emerging at $\mu=0.1$ (solid line). The horizontal dashed line shows the amplitude of the time averaged $Q/I$ profile.}
\label{stepan:fig:dyn}
\end{figure}

We have used the hydrodynamical simulation referred to as the strongly dynamic case by \citet{asensioramos03} and we have solved the H$\alpha$ scattering polarization problem in each 1D model corresponding to each time step of the hydrodynamical simulation. The results can be found in Fig.~\ref{stepan:fig:dyn}. The left panel shows the time variation of the kinetic temperature and proton density (which is the main source of collisional depolarization of H$\alpha$) at the height where the line optical depth for a line of sight with $\mu=0.1$ is unity. One can identify in Fig.~\ref{stepan:fig:dyn} the shock waves going through the atmospheric region. These are followed by a slow recombination of hydrogen. In the right panel of the same figure, the amplitude of the emergent $Q/I$ profile at $\mu=0.1$ is plotted as a function of time. Note that after the wave passes through the line formation region (which abruptly decreases the polarization signal), the emergent linear polarization slowly increases until the next shock arrives. These time intervals of increasing amplitudes correspond to the compression phases of the simulated chromospheric plasma. The $Q/I$ profiles with the largest amplitudes are thus red-shifted in the spectrum while the blue-shifted phases correspond to the smallest signals following immediately the shock wave passing. Interestingly, the average $Q/I$ profile, whose amplitude is very similar to the one obtained in the non-magnetic FAL-C model \citep[see Fig.~2 of][]{stepanjtb10asym}, does not show any noteworthy asymmetry at all.

\section{Spectropolarimetric observations and modeling}

The suggestion by \citet{stepanjtb10asym} of the existence of an abrupt change in the degree of magnetization of the upper chromosphere of the quiet Sun was based on 1D radiative transfer modeling of the peculiar line core asymmetry of the $Q/I$ profile observed by \citet{gandorfer00}, using semi-empirical models of the solar atmosphere such as the FAL-C model. Given that the real solar chromosphere is highly inhomogeneous and dynamic any such 1D static model can only be considered as a poor representation of the chromospheric thermal and density stratification. Therefore, we cannot exclude the possibility of an alternative explanation. Nevertheless, it is intereting to note that the temporally averaged $Q/I$ profile calculated without magnetic fields in the hydrodynamical simulations of chromospheric dynamics of \citet{carlssonstein97} does not show any noteworthy LCA (see Section~\ref{stepan:ssec:hydro}).

\begin{figure}
\includegraphics[width=5in]{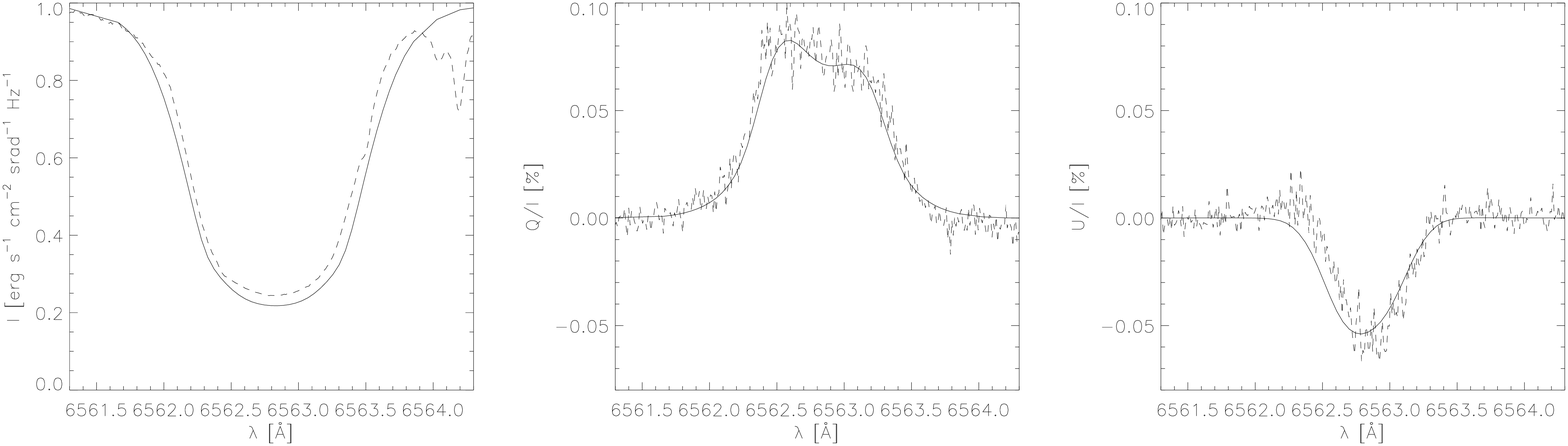}
\caption{An example of our recent 
spectropolarimetric observations of the H$\alpha$ Stokes profiles at
  about 8 arcsec inside the
solar limb (dashed lines)  obtained with ZIMPOL at IRSOL and the theoretical fit (solid lines) we have obtained by assuming an abrupt magnetization and an enhanced microturbulent velocity in the upper chromosphere of the FAL-C model.}
\label{stepan:fig:new}
\end{figure}

There are two inter-related avenues to confirm or refute the above-mentioned
tentative conclusion by \citet{stepanjtb10asym}. On the one hand, it is
important to develop realistic three-dimensional MHD simulations of the quiet
solar chromosphere, and to use them to compute the emergent $Q/I$ profile for
comparison with the  observations of \citet{gandorfer00}.  On the other hand,
it is crucial to carry out full Stokes observations of the H$\alpha$ line at
various on-disk positions. We are presently working on both fronts, but here
we focus only on showing an interesting example of one of our recent
spectropolarimetric observations carried out with ZIMPOL
\citep{gandorfer04} at the GCT of IRSOL \citep{bianda09},
which in addition to showing a similar LCA in the observed $Q/I$ profile it
shows a clear non-zero $U/I$ signal (see the dotted lines of
Fig.~\ref{stepan:fig:new}). 
The observations were obtained near the heliographic
  North pole at $\mu=0.13 \pm 0.01$ on June 13th 2009. The spectrograph slit
  width was set to 0.5 arcsec and the total exposure time was 800 seconds. The
  data reduction included the correction for instrumental polarization
  described by \citet{ramelli_2006}. The $Q/I$ and $U/I$ images show
  interesting structures and variability along the spatial direciton.
  The profiles reported here have been obtained integrating 16 arcsec along the spectrograph slit.
The solid lines of Fig.~\ref{stepan:fig:new} show a good fit to these new observations, which we have obtained by using the FAL-C model assuming an abrupt magnetization above a height 
 of 1900 km (i.e., similar to that of Fig. 3A in \citet{stepanjtb10asym}) and a fixed magnetic field inclination ($\theta_B{\approx}65^\circ$) and azimuth ($\chi_B{\approx}135^{\circ}$). It is however important to mention that although the core of $Q/I$ is well fitted when using the height-dependent microturbulent velocity of the FAL-C model, we had to artificially increase it up to about 20\,km\,s$^{-1}$ in order to be able to achieve a better fit of the profile wings. We interpret this as an indication of broadening by unresolved motions and/or structures in the real solar chromosphere.


\section{Conclusions and Perspectives}

The scattering polarization of the H$\alpha$ line
and its sensitivity (via the Hanle effect) to
magnetic fields between 0 and 50\,G is a promising tool for diagnostics of chromospheric magnetic fields. In this paper, we have described the peculiar character of the fine structure components of the H$\alpha$ line and of the formation of its scattering polarization profiles in optically thick model atmospheres, including the time-dependent hydrodynamical models of \citet{carlssonstein97}. Moreover, we have presented an interesting example of the full Stokes vector observations we are carrying out, which shows a conspicuous LCA in $Q/I$ and a clear $U/I$ signal.

For future progress, new observations of the H$\alpha$ polarization with higher spatial resolution within and outside coronal holes are urgently needed. The modeling of such observations with the help of the new generation of chromospheric models that are currently under development will facilitate new significant advances in the exploration of chromospheric magnetism via spectropolarimetry.


\acknowledgements  We are grateful to Mats Carlsson for providing us their hydrodynamical models of chromospheric dynamics. Financial support by the Spanish Ministry of Science and Innovation (project AYA2007-63881) and from the SOLAIRE network (MTRN-CT-2006-035484) is gratefully acknowledged.


\end{document}